\documentclass[aps,onecolumn,showpacs,superscriptaddress,preprintnumbers,amsmath,amssymb]{revtex4}
\usepackage{graphicx}
\usepackage{dcolumn}
\usepackage{bm}
\usepackage{color}
\usepackage{ulem}

\begin{document}

\title{Site-specific Interrogation of an Ionic Chiral Fragment During Photolysis Using an X-ray Free-Electron Laser}

\author{\firstname{Markus} \surname{Ilchen}}\email{markus.ilchen@xfel.eu}
\affiliation{Institut f\"{u}r Physik und CINSaT, Universit\"{a}t Kassel, Heinrich-Plett-Stra\ss e~40, D-34132 Kassel, Germany}
\affiliation{European XFEL GmbH, Holzkoppel 4, 22869 Schenefeld, Germany}
\affiliation{Stanford PULSE Institute, 2575 Sand Hill Road, Menlo Park, California 94025, USA}

\author{\firstname{Philipp} \surname{Schmidt}}
\affiliation{Institut f\"{u}r Physik und CINSaT, Universit\"{a}t Kassel, Heinrich-Plett-Stra\ss e~40, D-34132 Kassel, Germany}
\affiliation{European XFEL GmbH, Holzkoppel 4, 22869 Schenefeld, Germany}

\author{\firstname{Nikolay M.} \surname{Novikovskiy}}
\affiliation{Institut f\"{u}r Physik und CINSaT, Universit\"{a}t Kassel, Heinrich-Plett-Stra\ss e~40, D-34132 Kassel, Germany}
\affiliation{Institute of Physics, Southern Federal University, 344090 Rostov-on-Don, Russia}

\author{\firstname{Gregor} \surname{Hartmann}}
\affiliation{Institut f\"{u}r Physik und CINSaT, Universit\"{a}t Kassel, Heinrich-Plett-Stra\ss e~40, D-34132 Kassel, Germany}
\affiliation{Helmholtz-Zentrum Berlin f\"{u}r Materialien und Energie, Albert-Einstein-Stra\ss e 15, 12489 Berlin, Germany}

\author{\firstname{Patrick} \surname{Rupprecht}}
\affiliation{Max-Planck-Institut f\"{u}r Kernphysik Heidelberg, Saupfercheckweg 1, 69117 Heidelberg, Germany}

\author{\firstname{Ryan N.} \surname{Coffee}}
\affiliation{SLAC National Accelerator Laboratory, 2575 Sand Hill Road, Menlo Park, California 94025, USA}

\author{\firstname{Arno} \surname{Ehresmann}}
\affiliation{Institut f\"{u}r Physik und CINSaT, Universit\"{a}t Kassel, Heinrich-Plett-Stra\ss e~40, D-34132 Kassel, Germany}

\author{\firstname{Andreas} \surname{Galler}}
\affiliation{European XFEL GmbH, Holzkoppel 4, 22869 Schenefeld, Germany}

\author{\firstname{Nick} \surname{Hartmann}}
\affiliation{SLAC National Accelerator Laboratory, 2575 Sand Hill Road, Menlo Park, California 94025, USA}

\author{\firstname{Wolfram} \surname{Helml}}
\affiliation{Technische Universit\"{a}t Dortmund, Fakult\"{a}t f\"{u}r Physik, Maria-G\"{o}ppert-Mayer-Stra\ss e, 44227 Dortmund, Germany}

\author{\firstname{Zhirong} \surname{Huang}}
\affiliation{SLAC National Accelerator Laboratory, 2575 Sand Hill Road, Menlo Park, California 94025, USA}

\author{\firstname{Ludger} \surname{Inhester}}
\affiliation{Center for Free-Electron Laser Science (CFEL), Notkestra\ss e 85, 22607 Hamburg, Germany}
\affiliation{Deutsches Elektronen-Synchrotron DESY, Notkestra\ss e 85, 22607 Hamburg, Germany}

\author{\firstname{Alberto A.} \surname{Lutman}}
\affiliation{SLAC National Accelerator Laboratory, 2575 Sand Hill Road, Menlo Park, California 94025, USA}

\author{\firstname{James P.} \surname{MacArthur}}
\affiliation{SLAC National Accelerator Laboratory, 2575 Sand Hill Road, Menlo Park, California 94025, USA}

\author{\firstname{Timothy} \surname{Maxwell}}
\affiliation{SLAC National Accelerator Laboratory, 2575 Sand Hill Road, Menlo Park, California 94025, USA}

\author{\firstname{Michael} \surname{Meyer}}
\affiliation{European XFEL GmbH, Holzkoppel 4, 22869 Schenefeld, Germany}

\author{\firstname{Valerija} \surname{Music}}
\affiliation{Institut f\"{u}r Physik und CINSaT, Universit\"{a}t Kassel, Heinrich-Plett-Stra\ss e~40, D-34132 Kassel, Germany}
\affiliation{European XFEL GmbH, Holzkoppel 4, 22869 Schenefeld, Germany}

\author{\firstname{Heinz-Dieter} \surname{Nuhn}}
\affiliation{SLAC National Accelerator Laboratory, 2575 Sand Hill Road, Menlo Park, California 94025, USA}

\author{\firstname{Timur} \surname{Osipov}}
\affiliation{SLAC National Accelerator Laboratory, 2575 Sand Hill Road, Menlo Park, California 94025, USA}

\author{\firstname{Dipanwita} \surname{Ray}}
\affiliation{SLAC National Accelerator Laboratory, 2575 Sand Hill Road, Menlo Park, California 94025, USA}

\author{\firstname{Thomas J. A.} \surname{Wolf}}
\affiliation{Stanford PULSE Institute, 2575 Sand Hill Road, Menlo Park, California 94025, USA}
\affiliation{SLAC National Accelerator Laboratory, 2575 Sand Hill Road, Menlo Park, California 94025, USA}

\author{\firstname{Sadia} \surname{Bari}}
\affiliation{Deutsches Elektronen-Synchrotron DESY, Notkestra\ss e 85, 22607 Hamburg, Germany}

\author{\firstname{Peter} \surname{Walter}}
\affiliation{SLAC National Accelerator Laboratory, 2575 Sand Hill Road, Menlo Park, California 94025, USA}

\author{\firstname{Zheng} \surname{Li}}
\affiliation{Center for Free-Electron Laser Science (CFEL), Notkestra\ss e 85, 22607 Hamburg, Germany}
\affiliation{State Key Laboratory for Mesoscopic Physics, School of Physics, Peking University, Beijing 100871, China}

\author{\firstname{Stefan} \surname{Moeller}}
\affiliation{SLAC National Accelerator Laboratory, 2575 Sand Hill Road, Menlo Park, California 94025, USA}

\author{\firstname{Andr\'e} \surname{Knie}}
\affiliation{Institut f\"{u}r Physik und CINSaT, Universit\"{a}t Kassel, Heinrich-Plett-Stra\ss e~40, D-34132 Kassel, Germany}

\author{\firstname{Philipp V.} \surname{Demekhin}}\email{demekhin@physik.uni-kassel.de}
\affiliation{Institut f\"{u}r Physik und CINSaT, Universit\"{a}t Kassel, Heinrich-Plett-Stra\ss e~40, D-34132 Kassel, Germany}

\date{\today}

\begin{abstract}
Short-wavelength free-electron lasers with their ultrashort pulses at high intensities have originated new approaches for tracking molecular dynamics from the vista of specific sites. X-ray pump X-ray probe schemes even allow to address individual atomic constituents with a ‘trigger’-event that preludes the subsequent molecular dynamics while being able to selectively probe the evolving structure with a time-delayed second X-ray pulse. Here, we use a linearly polarized X-ray photon to trigger the photolysis of a prototypical chiral molecule, namely trifluoromethyloxirane (C$_3$H$_3$F$_3$O), at the fluorine K-edge at around 700~eV. The evolving fluorine-containing fragments are then probed by a second, circularly polarized X-ray pulse of higher photon energy in order to investigate the chemically shifted inner-shell electrons of the ionic motherfragment for their stereochemical sensitivity. We experimentally demonstrate and theoretically support how two-color X-ray pump X-ray probe experiments with polarization control enable XFELs as tools for chiral recognition.
\end{abstract}

%\keywords{Photoionization of atoms and ions, multiphoton ionization and excitation to highly excited ions, X-ray spectra, Optical activity and dichroism, Chiral media}

\maketitle

\newpage 

%\section{Introduction}
%\label{sec:intro}	

Chirality is a fundamental phenomenon that determines everybody's all-day-life to a large extent. It is, for instance, responsible for odor perception, taste, and the effect of most pharmaceutical drugs on any living organism. In fact, the molecular building blocks of all known biological matter possess a handedness. They are non-superimposable mirror images of themselves, called enantiomers. It currently constitutes an interdisciplinary effort to find the underlying reason for this and its origin. Determining and controlling the conformation of chiral molecules is not only an important aspect of the natural sciences but also an industrial, specifically, pharmaceutical research subject. However, chiral molecules are identical in their stoichiometric composition and they are practically only distinguishable via their interaction with other chiral objects.\\

A versatile tool for investigating enantiomers in the gas phase in a controlled and theoretically well accessible way is circularly polarized light, since it possesses a handedness as well. The so-called circular dichroism (CD) quantifies the absorption difference between left and right circularly polarized light and is typically an effect accounting only for a few per mille of the overall signal. This light--matter interaction is today commonly used for chiral recognition and control. In 1976, Ritchie \cite{Ritchie76} predicted that circularly polarized light can imprint its enantio-selectivity onto emitted  photoelectrons. His groundbreaking idea can be visualized with a picture of an electron driven by circularly polarized light which scatters on the asymmetric potential of a chiral molecule like a nut on a thread, meaning it either moves in forward or in backward direction with respect to the propagation of the light, depending on the rotational direction of the components. This so-called photoelectron circular dichroism (PECD) can be orders of magnitude stronger than the respective absorption difference of circularly polarized photons, since it emerges already in the electric-dipole approximation. \\

At present, PECD has been exploited for chiral recognition using a variety of light sources and photon energy ranges. Low photon energy has, for example, been used for directly exploring the chiral chemical environment \cite{Nahon15,Hadidi18}, and higher photon energies to address individual constituents of a molecule and to explore how specific sites encounter the surrounding chiral potential \cite{Hergenhahn04,Ilchen17}. Contrary to the case of outer-shell photoionization with low photon energies where both, initial and final electronic states contribute to the observed effect, the case of high photon energy photoionization from an almost symmetric atomic-like inner shell allows one to study an individual contribution of the final electron continuum state. Furthermore, strong laser sources have been used to explore how a chiral molecule interacts with multiple photon absorptions in the long-wavelength regime \cite{Lux12,Lehmann13}. Finally, time-resolved investigations have recently started to explore ultrafast processes in chiral compounds with femtosecond optical laser pulses \cite{Beaulieu16,Comby16}. The versatility of PECD is valuable but also a challenge to the differentiation of individual properties and processes of chirality \cite{Wollenhauptreview}.\\

Today's short-wavelength FELs provide photon pulses of very high brilliance, i.e.\ they can exceed the brilliance of any other light source at the accessible photon energies by more than a billion times. The high intensity allows for very efficient time-resolved experiments that either use optical lasers in order to trigger processes in the target of choice \cite{optXray} or even X-ray pulses as so called `pump' \cite{XrayXray1, XrayXray2, XrayXray3, XrayXray4, XrayXray5, XrayXray6}, while in both cases, the dynamics are probed with a second X-ray pulse. Their capability of enabling high-precision studies of dichroic light--matter interaction in dilute species was initially demonstrated in the vacuum ultraviolet regime at FERMI in Italy \cite{Allaria14, Mazza14}. The first circularly polarized X-ray pulses that enabled high irradiation levels were achieved at the LCLS \cite{Lutman16, Hartmann16}. Such polarization-controlled (X)FELs naturally offer the possibility for ultrafast studies of chiroptically sensitive targets from the perspective of element- and even site-specific constituents during ultrafast structural changes or electron migration.\\

The present work reports the first investigation of a site-specific PECD in an ionic chiral fragment of a dissociating chiral molecule by an XFEL. The employed X-ray pump --  X-ray probe scheme allows for addressing specific atoms in the molecule in both pulses. In particular, an ultrashort X-ray pump pulse triggers photolysis of the prototypical molecule trifluoromethyloxirane (C$_3$H$_3$F$_3$O) in a way that, predominantly, one singly-charged fluorine atom dissociates and leaves the singly-charged mother-fragment behind. While the distance between these two ionic systems increases in time, a second ultrashort X-ray pulse probes the PECD of F 1s photoelectrons of the remaining mother-fragment. This possible channel of dissociation is one out of many, however, it can be energetically disentangled via electron VMI spectroscopy. The static PECD of F 1s photoelectrons of neutral trifluoromethyloxirane molecules was systematically studied both, experimentally and theoretically, in our previous work \cite{Ilchen17}. This preliminary work uncovers which chiral asymmetry can be expected in a broad interval of the photoelectron kinetic energies and demonstrates that the presently used theoretical approach can describe the measured PECD quantitatively. In the present work, we theoretically demonstrate that the PECD of a fluorine core-photoelectron emitted from the chiral mother-fragment can sense the dissociating fluorine ion over a wide range of internuclear distances, far beyond chemical bond dynamics. Averaging these dynamics in the predicted stereochemical sensitivity of the mother-fragment over the individual fluorine ejection and ionization channels, surprisingly, results in a slightly oscillating PECD of around 1\% to 2$\%$. This finding is qualitatively supported by the experimental data.

\section{Results and discussion}
\label{sec:Results}	

\subsection{The process}
\label{sec:Process}

\begin{figure}[b!]
\includegraphics[width=0.5\columnwidth]{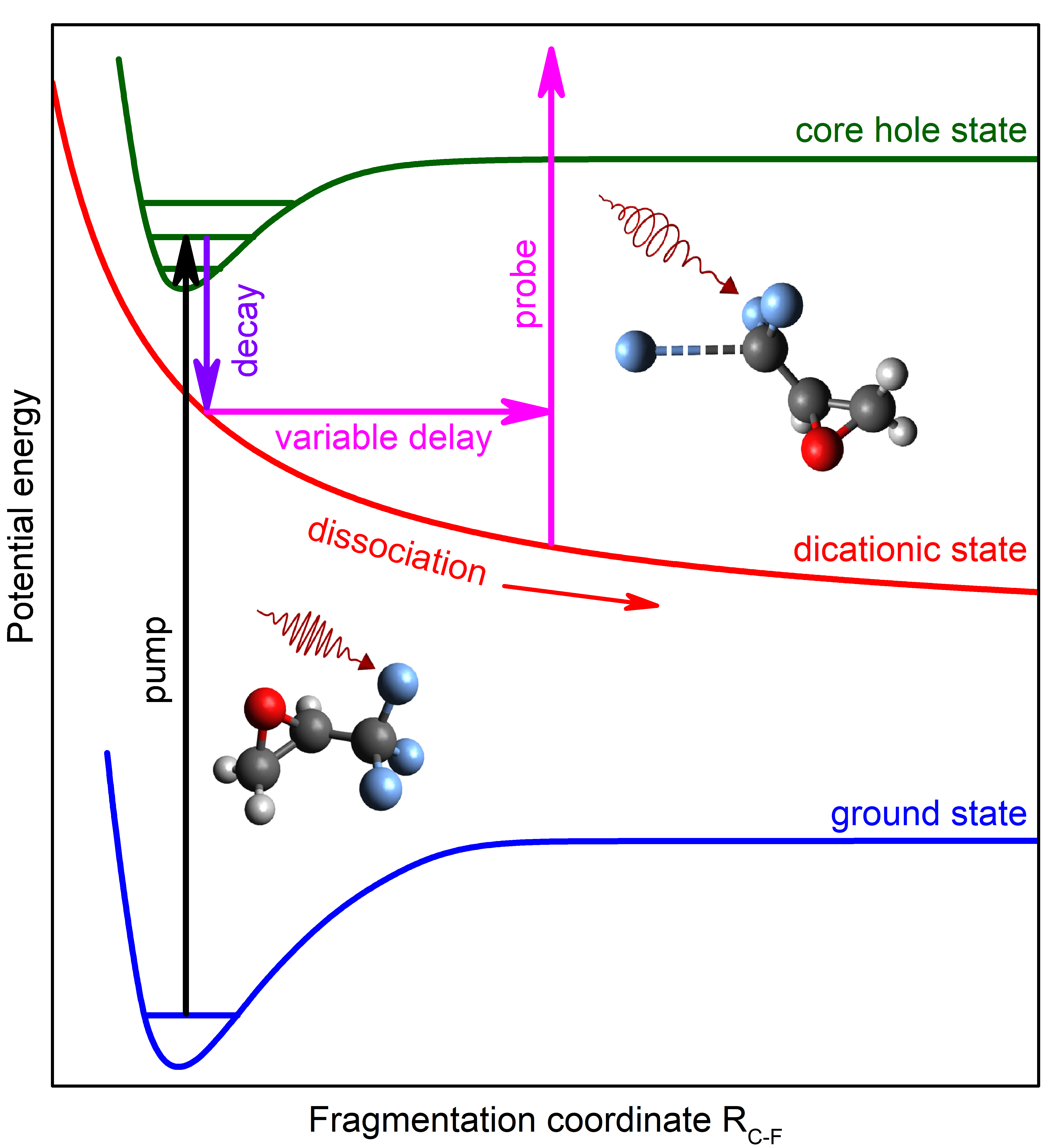}
\caption{Principal scheme of the process. The X-ray pump pulse ejects a 1s photoelectron from one of the fluorine atoms of the neutral C$_3$H$_3$F$_3$O molecule. This step is represented by the upward transition from the ground state to the core-ionized state of the molecule. In the second step, the core-ionized state undergoes Auger decay which populates dicationic states of the molecule as indicated by the downward transition pointing to the repulsive curve. The doubly-charged molecule can Coulomb-explode into the fluorine ion and the singly-charged mother-fragment. During this dissociation, a delayed, circularly polarized X-ray pulse probes the asymmetric scattering of the 1s photoelectron emitted from one of the remaining fluorine atoms attached to the singly-charged mother-fragment. This second electron emission in an altered Coulomb potential of the transient system is shown by the upward transition at an enlarged distance between the ionic fragments. 
The present experiment analyzes the PECD of this secondary F 1s photoelectron of C$_3$H$_3$F$_2$O$^+$ as a function of the time delay between the pulses.}\label{fig:scheme}
\end{figure}

The presently studied process is schematically illustrated in Fig.~\ref{fig:scheme}. The X-ray pump pulse with a photon energy of about 698~eV releases a 1s photoelectron from one of the fluorine atoms in the neutral C$_3$H$_3$F$_3$O molecule with a kinetic energy of about 4~eV. This corresponds to a calculated F 1s binding energy of about 694~eV. In the next step, the created core-ionized state decays by an ultrafast Auger process (the presently calculated lifetime is about 2.3~fs) and populates dicationic states in the molecule. The F 1s binding energy of the doubly-charged but still intact molecule in the ground state is calculated to be about 704.7 eV. After charge rearrangement, the created dicationic states initiate a Coulomb explosion, predominantly into a singly-charged fluorine ion F$^+$, i.e.\ the one which was initially ionized, and the also singly-charged mother-fragment C$_3$H$_3$F$_2$O$^+$. The F 1s binding energy of the mother-fragment, after having undergone the full chemical shift due to a missing fluorine ion, is estimated to be about 8~eV higher compared to that in the neutral molecule, i.e.\ about 702~eV. Previous studies support that this channel under investigation can indeed be expected to be prominent \cite{Iwayama13}.\\

In the course of this fragmentation, a delayed, circularly-polarized X-ray pulse with a photon energy of about 710~eV ionizes the 1s electron of one of the fluorine atoms in the C$_3$H$_3$F$_2$O$^+$ at different distances to the ejected F$^+$ ion. A second photoelectron with a kinetic energy of about 8~eV is emitted. Those secondary photoelectrons can be energetically separated from the primary photoelectrons and from other nonlinearly generated electronic contributions. In fact, signals of the photoelectrons released from the charged system by the probe pulse scale nonlinearly with the X-ray pulse energy, i.e.\ quadratic for a two-photon absorption. Therefore, they can be discriminated from the photoelectrons released from the neutral molecule, whose signals scale linearly with the XFEL pulse intensity. Details on the data analysis and a discussion of alternative possible processes to be taken into account can be found in  Sec.~\ref{sec:methodsDATA}. In the present experiment, the forward--backward asymmetry, i.e.\ PECD, of the secondary F 1s photoelectrons released from the singly-charged mother fragment C$_3$H$_3$F$_2$O$^+$ by the probe pulse is analyzed, while the initially ionized fluorine atom increases its distance to the remaining chiral structure along the respective C--F bond.

\subsection{Theoretical estimates}
\label{sec:ResultsTHEO}

For enabling access to the transient PECD of the F 1s photoelectrons ejected from the singly-charged mother-fragment, we modeled the final step of the process theoretically. It should be stressed, that a full theoretical treatment of even this last photoionization step is currently almost impossible. This is because the Auger decay of the initially core-ionized  F$^+$(1s$^{-1}$) in the trifluoromethyloxirane molecule populates a huge manifold of different dicationic states, which leave the created F$^+$ and C$_3$H$_3$F$_2$O$^+$ fragments in a variety of different electronic configurations. As a consequence, different channels generate different chiral potentials for the scattering of the secondary F 1s photoelectron. In addition, the internuclear geometry of the singly-charged mother-fragment relaxes dynamically in the course of the fragmentation. Moreover, since the center of charge of the C$_3$H$_3$F$_2$O$^+$ fragment does not necessarily coincide with the center of its mass, the mother-fragment may rotate as a whole, alternating thereby the relative geometry of the two species.\\

Because of these difficulties, we made a qualitative estimate of what can be expected as outcome of such an experiment. In particular, the present theoretical model implies the following restrictions. First of all, we used a frozen internuclear geometry at the equilibrium of the neutral molecule in its ground state. Secondly, we assumed straight dissociation of the F$^+$ ion along the respective C--F bond, without possible changes in the relative orientation of the ions. Finally, we assumed that both, the  F$^+$ and  C$_3$H$_3$F$_2$O$^+$ ions end up in their electronic ground configurations after Auger decay. Under those assumptions, the PECDs of the 1s photoelectrons of individual fluorine atoms remaining bound to the mother-fragment were computed for the second ionization step alone at different separations to the dissociating F$^+$ ion in the photoelectron energy range of 6--10~eV. These calculations were performed by the single center (SC) method and code \cite{SC1,SC2}, as outlined in Sec.~\ref{sec:methodsTHEORY}. \\

\begin{figure}
\includegraphics[width=0.5\columnwidth]{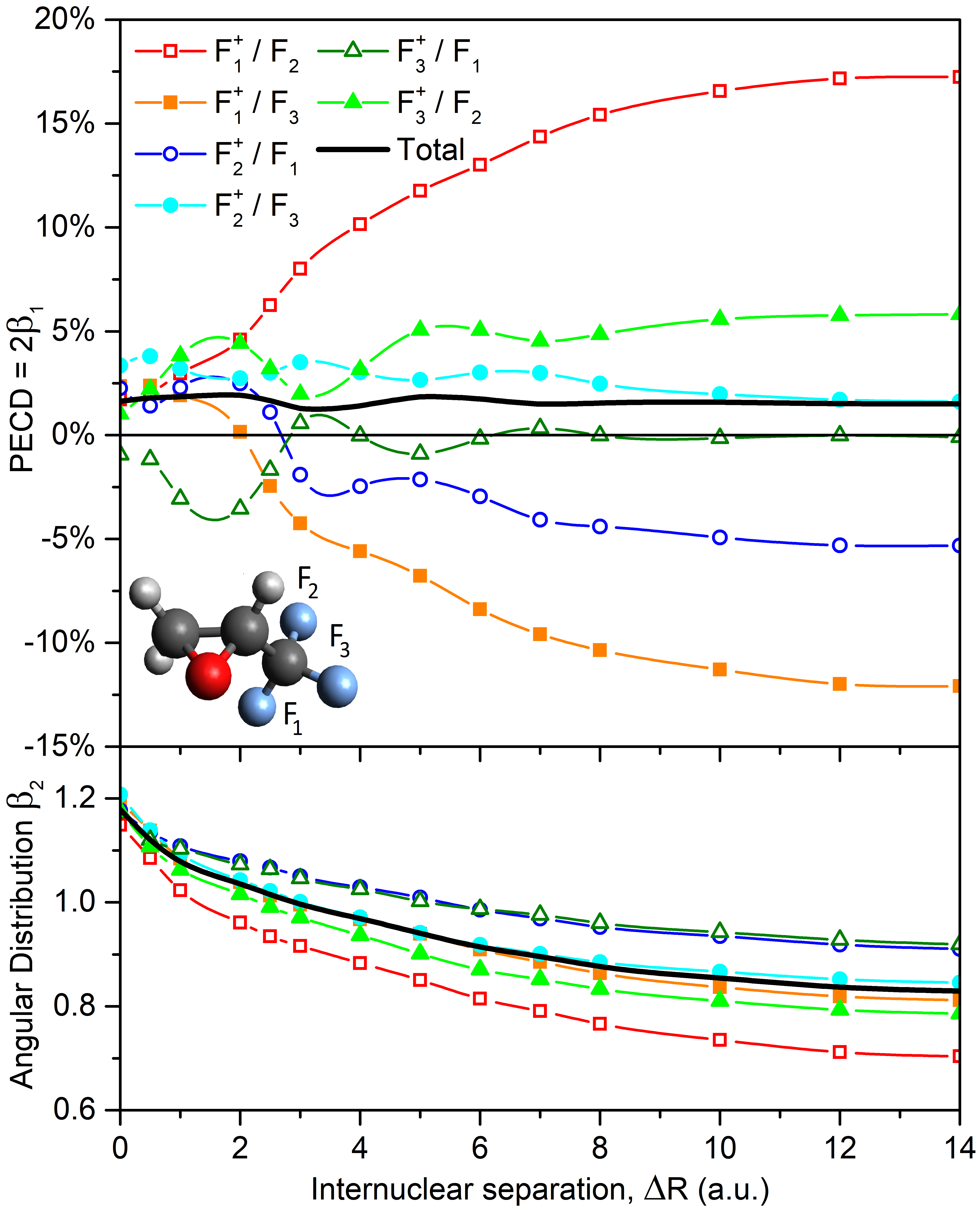}
\caption{PECD (upper panel) and angular distribution parameter (lower panel) of 1s photoelectrons of the individual fluorine atoms F$_\mathrm{i}$ (as enumerated in the inset-picture) remaining bound to the singly-charged C$_3$H$_3$F$_2$O$^+$ mother-fragment and the respective averaged quantities, computed for the R-trifluoromethyloxirane at different distances $R_{\mathrm{C-F}}=R_{\mathrm{eq}}+\Delta R$ between the C atom of the CF$_2$-group and the individual  F$^+_\mathrm{j}$ fluorine ions dissociating  along the bond (see legend for details).}\label{fig:theory}
\end{figure}

The present theoretical results are collected in Fig.~\ref{fig:theory}. This figure depicts the PECD (upper panel) as twice the dichroic parameter $\beta_1$ and the angular distribution parameter $\beta_2$ (lower panel), computed for all possible combinations of the dissociating fluorine ion F$^+_\mathrm{j}$ and the ionized fluorine atom F$_\mathrm{i}$, as a function of the internuclear separation $R_{\mathrm{C-F}}=R_{\mathrm{eq}}+\Delta R$. Each depicted point represents a value averaged over the studied kinetic energy interval to account for the experimental photon energy bandwidth of $\approx 3.5$~eV. As one can see for the combination of the F$_2$ atom and the F$_1^+$ ion, the computed PECD grows almost monotonically from a small value of about +2\% at $R_{\mathrm{C-F}}=R_{\mathrm{eq}}$, i.e.\ at $\Delta R=0$, to a large value of about +17\%  at $\Delta R=14$~a.u. For the second remaining fluorine atom F$_3$ and the same dissociating ion F$_1^+$, the computed PECD exhibits an opposite behaviour, i.e.\ it monotonically decreases from a similar positive value of about +2\% at $\Delta R=0$ to a large negative value of about --12\% at $\Delta R=14$~a.u., changing thereby its sign around $\Delta R=2$~a.u. \\

For the dissociation of the second F$_2^+$ ion and ionization of the two remaining atoms F$_1$ and F$_3$, as well as of the third F$_3^+$ ion and the F$_1$ and F$_2$ atoms, changes of the computed PECDs as functions of the internuclear separation $\Delta R$ are not as strong, i.e.\ well within $\pm6$\%. Those PECDs possess distinct oscillations, which are comparable to the whole variation range, and can change their sign (even multiply as can be seen for the F$_3^+$/F$_1$-combination). Such oscillations might be a result of interference effects in the multiple scattering of the photoelectron waves emitted from the mother-fragment on  the dissociating fluorine ions. It is interesting to note that, for a given dissociating F$^+_\mathrm{j}$ ion, PECDs of the two remaining F$_\mathrm{i}$ atoms exhibit qualitatively `opposite' trends. On the contrary, the computed individual angular distribution parameters $\beta_2$ possess very similar trends: They fall monotonically with an increase of the internuclear separation from a value in between 1.15--1.21 to a value in between 0.70--0.92. As can also be seen from Fig.~\ref{fig:theory}, the computed individual PECDs and $\beta_2$-parameters are almost saturated to their asymptotic (non-zero) values at the largest considered internuclear distance of $\Delta R=14$~a.u.

\newpage

The black solid curves in Fig.~\ref{fig:theory} represents the total PECD and $\beta_2$-parameter, averaged over all individual contributions, since photoelectrons released in different F$_\mathrm{j}^+$/F$_\mathrm{i}$-combinations cannot be distinguished in the present experiment. Because of almost opposite behaviors of the chiral asymmetries computed for individual combinations, their strong changes and oscillations as functions of the internuclear separation cancel out, resulting in a moderate total chiral asymmetry which varies in between  +1.2\%  and +1.9\%. Although these theoretical results should be considered as a qualitative estimate of the effect, Fig.~\ref{fig:theory} allows to draw a few important conclusions. For the considered photoelectron kinetic energy interval, the PECD computed for the individual F$_\mathrm{j}^+$/F$_\mathrm{i}$-combinations  can change as a function of the internuclear separation between the two fragments  by almost an order of magnitude. Those individual chiral asymmetries can distinctly oscillate with the internuclear separation, and can even change their sign. The described variations and changes persist for a rather large internuclear separation beyond the chemical bond breaking, where the $R_{\mathrm{C-F}}$ bond is extended beyond $\Delta R=10$~a.u. The latter  observation can be related to the long-range impact of the Coulomb repulsion between the two ionic fragments.

\subsection{Experimental results and discussion}
\label{sec:ResultsEXPT}	

The present experiment was carried out at the AMO beamline of the LCLS at the SLAC National Accelerator Laboratory in the USA \cite{LCLS_AMO}. Both intense pump and probe X-ray pulses had a similar duration of about 10$\pm$3~fs. To be able to energetically distinguish the primary and secondary photoelectrons and to specifically address the F 1s electrons of the singly-charged C$_3$H$_3$F$_2$O$^+$ mother-fragment at an enantio-sensitive kinetic energy, the photon energies of the pump and probe pulses were set to 698$\pm$2~eV and 710$\pm$2~eV, respectively. In order to minimize a possible influence of the pump pulse on the studied PECD effect, we used a linearly polarized pump pulse, while the time-delayed probe pulse was circularly polarized. The emitted photoelectrons were detected with a VMI spectrometer, which provided access to their laboratory-frame angular emission patterns. The measurements were performed for four time delays between the pulses at $\tau$=0, 60, 125, and 250~fs. More details on the beamline operation, experimental setup, and data analysis are provided in the Secs.~\ref{sec:methodsLCLS}--\ref{sec:methodsDATA}.\\

In the obtained VMI-spectrum, we observe different photoelectron signals. The photoelectrons from the neutral molecules ejected by the linearly-polarized pump pulse emerge in the low kinetic energy range at around 4~eV and cannot exhibit any chiral asymmetry. Photoelectrons released by the circularly-polarized probe pulse from the neutral molecules, at around 16~eV, are not found to possess any significant PECD, presumably because of their rather high kinetic energy. These observations are consistent with our previous study of the F 1s photoionization of trifluoromethyloxirane \cite{Ilchen17}. As also expected, photoelectrons released by the circularly-polarized probe pulse from the neutrally ejected fluorine atoms, at around 12.5~eV kinetic energy, exhibit no chiral asymmetry. On the contrary, we observe an enantio-selective chiral asymmetry for the secondary F 1s photoelectrons from the singly-charged mother-fragment C$_3$H$_3$F$_2$O$^+$ with a kinetic energy of 8~eV. For the time delay $\tau$=0~fs, i.e. when the two pulses overlap, a variety of processes occurs within the duration of the pulses, such as Auger decay, charge-migration dynamics, restructuring before the actual bond-breaking, and rapid change of the Coulomb potential. All these processes significantly alter the observed F 1s photoelectron signal from C$_3$H$_3$F$_2$O$^+$, and its respective PECD which can, therefore, not be decisively extracted. For longer delays, when most ultrafast relaxation processes are essentially completed, the present data analysis could be performed in a more consistent way. In particular, for the time delays of $\tau$=60, 125 and 250~fs and the R-enantiomer, we observe PECDs of $+2.7\%(+1.0\%/-2.0\%)$, $+0.3\%(\pm2.6\%)$, and $+0.9\%(\pm3.0\%)$, respectively. The PECD of $-2.5\%(-1.0\%/+2.0\%)$ measured for the S-enantiomer and $\tau$=60~fs time delay confirms a chiral origin of the observed asymmetry.\\

\begin{figure}
\includegraphics[width=0.5\columnwidth]{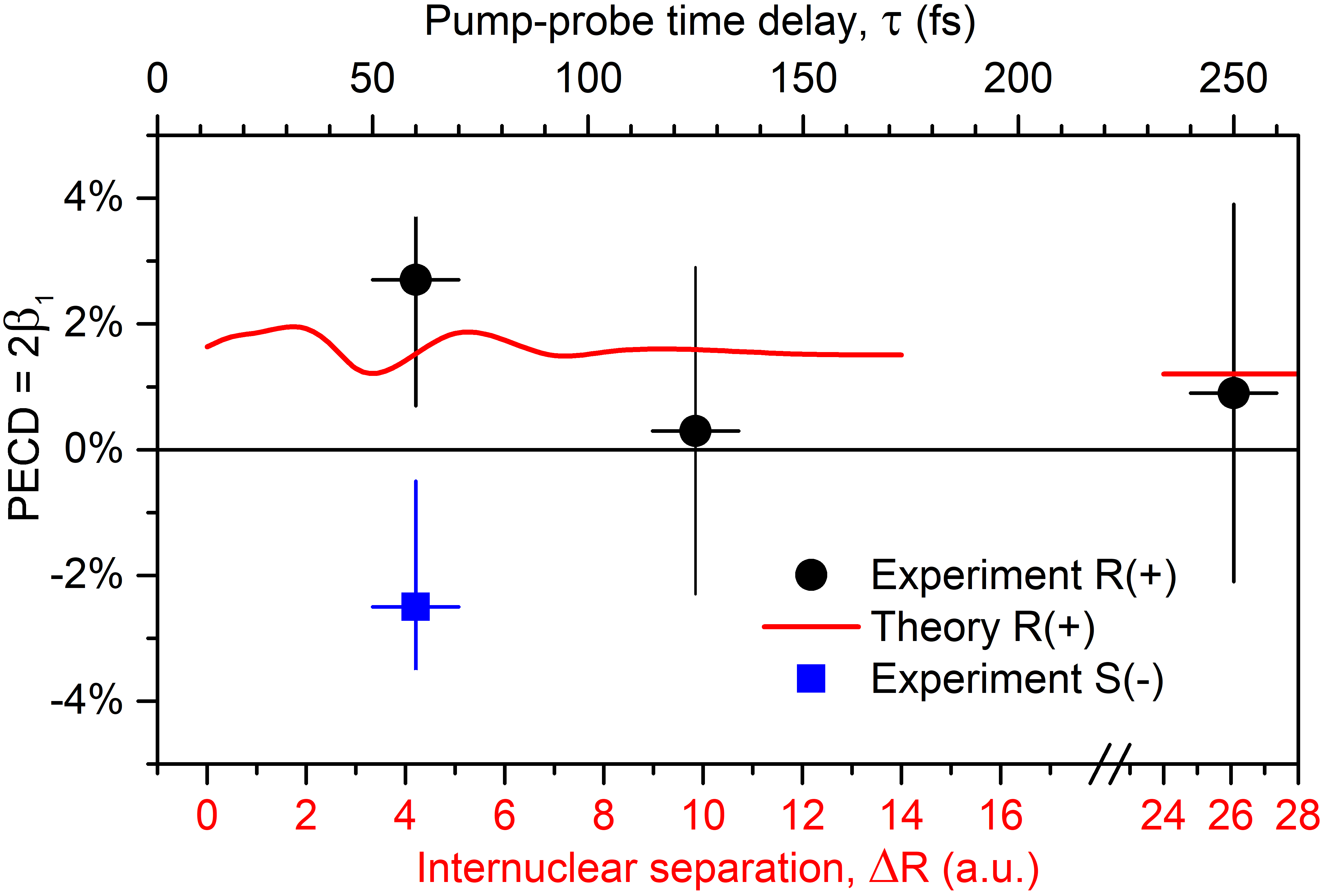}
\caption{Average PECD of all combinations of remaining and dissociating fluorine atoms in singly-charged C$_3$H$_3$F$_2$O$^+$ in its R-enantiomeric form, measured at different time delays between the pump and probe X-ray pulses (shown by circles with error bars; refers to the upper horizontal scale), together with the corresponding average theoretical PECD (solid curve), computed at different internuclear separations $R_{\mathrm{C-F}}=R_{\mathrm{eq}}+\Delta R$ between the C atom on the CF$_2$-group of the mother-fragment and the dissociating F$^+$ ion along the respective C--F bond (refers to the lower horizontal scale). The two horizontal scales are adjusted via a classical Coulomb explosion model (see text for details). The theoretical value at the largest shown separations represents an asymptotic PECD for $R_{\mathrm{C-F}}=\infty$. Experimental PECD measured for S-enantiomer at the delay of 60~fs is shown by solid square with error bars (refers to the upper horizontal scale).
} \label{fig:results}
\end{figure}

These experimental results are collected in Fig.~\ref{fig:results} as a function of the time delay (upper horizontal scale). Owing to the large experimental uncertainties (see Sec.~\ref{sec:methodsDATA} for details), an unambiguous PECD can only be observed for the time delay of $\tau$=60~fs. The red curve in Fig.~\ref{fig:results} depicts the average theoretical PECD from Fig.~\ref{fig:theory} (black solid curve) as a function of the internuclear separation $R_{\mathrm{C-F}}=R_{\mathrm{eq}}+\Delta R$  (lower  horizontal scale; note also the axis breaking). In order to interrelate the internuclear separation with the pump-probe time delay, we used a simplified classical Coulomb-explosion model. In the model, we considered two point charges, one localized on the dissociating fluorine and another on the oxirane ring, being initially separated by 5.5~a.u. The model yields, that the considered time delays of 60, 125, and 250~fs correspond to the internuclear separations of about $\Delta R \approx $ 4, 10 and 26~a.u. \\

It should be stressed, however, that the used Coulomb-explosion model is by far oversimplified. Indeed, as the F$^+$ fragment moves apart, the positive charge of the double-hole instantly delocalizes over the molecular skeleton in order to reduce the Coulomb-repulsion force. Owing to this dynamical charge delocalization and additional geometry relaxation, the realistic fragmentation dynamics does not follow an expected $1/R$ repulsion law. In addition, a part of the repulsion energy can be transferred to the remaining manifold of internal nuclear degrees of freedom of the whole system, and owing to this energy dissipation, the overall separation should proceed on a slower timescale. Nevertheless, since the calculated average chiral asymmetry changes only moderately over the whole internuclear distance range, we rely in the further discussion on the implied Coulomb-explosion model. \\

As one can see from Fig.~\ref{fig:results}, the average theoretical and experimental PECDs exhibit a similar value for $\tau$=60~fs and $\Delta R \approx 4$~a.u., whereas the measured asymmetry is somewhat stronger (note that the large experimental uncertainties do not allow for a more decisive interpretation). The computed  PECD becomes almost constant ($\approx$+1.5\%) at the internuclear separations of about $\Delta R \approx 10$~a.u. (corresponding to a time delay of about $\tau$=125~fs). Further on, it saturates very slowly to its asymptotic value of about  +1.2\% (shown by the red horizontal bar at the internuclear separations of around $\Delta R \approx 26$~a.u. after the axis break). These theoretical predictions agree with the experimental results within the respective uncertainties, which reveals a PECD on the targeted singly-charged chiral fragment created via an X-ray photon. A decisive confirmation that the experimental PECD indeed drops down at 125~fs requires more accurate calculations of the complicated fragmentation dynamics as well as advanced experimental methods to reduce the present uncertainties.

\section{Conclusions and outlook}
\label{sec:Conclusions}	

In the present work, we fragment enantiomerically pure trifluoromethyloxirane and study the specific channel that creates a singly-charged chiral mother-fragment C$_3$H$_3$F$_2$O$^+$ and a singly-charged fluorine atom. The PECD of the F 1s photoelectrons of the ionic mother-fragment is the subject of the present investigation. We employ an X-ray pump -- X-ray probe scheme enabled by the LCLS-XFEL at SLAC in the USA. For the F 1s photoionization of C$_3$H$_3$F$_2$O$^+$ in the vicinity of a dissociating F$^+$ ion via a circularly polarized X-ray pulse, we theoretically predict a highly-dynamical chiral asymmetry that differs enormously for different emitter-site perspectives. The absolute values of the individual channels' PECD can grow as high as 17$\%$. It is remarkable that the PECD provides a dynamical screening of the chirality of the mother-fragment even at very large internuclear distances to the ejected fluorine ion. This indicates, that the chiral potential of the mother-fragment is still influenced by the presence of the dissociating fluorine ion,  similar to findings of a recent time-resolve study of an ordinary circular dichroism (CD) of a chiral molecule \cite{Woerner19}. The prediction of a persisting average PECD of about 1\% agrees qualitatively with the experimental results within the present experimental uncertainties and quantitatively for the time delay at $\tau$=60~fs.\\

It should be noted that the quality of the presently performed experiment to access time-resolved PECD is compromised by the complexity of the nuclear and electron dynamics accompanying the process. The advantages of the inner-shell ionization of the atomic fragment, which allows an unambiguous assignment of the origin of the secondary photoelectrons, can potentially be utilized in a more controlled way if e.g. (neutral) dissociation of a specific (unique) element is governing the fragmentation dynamics. This can in principle be achieved, e.g., by an optical laser pulse preceding an ionizing XFEL pulse. There are cases where both, a dissociating atom and its mother-fragment, could mainly stay in their neutral ground states and no significant charge and bond rearrangements, as well as rotations of the fragments with respect to each other would be expected. Very importantly, neutral dissociation evolves on a much slower time scale, enabling higher resolution in the time domain.\\

The present work is the first step on a road map that is set to pave the way for future investigations of chirality with free-electron lasers. Tackling the underlying challenges effectively requires robust understanding of ultrafast, nonlinear and polarization dependent physics that in several cases lies at the very limit of today's technical capabilities and insights. All of the mentioned topics entail independent and vital fields of state-of-the-art science. However, only in combination, a reasonably broad basis can be established to allow for the site-specific study of fundamental and applied dynamics in complex chiral systems. Besides further establishing the fundamentals of nonlinear and ultrafast light--matter interaction in chiral systems, future studies of dynamics in, e.g., partially chiral peptides and larger complexes, transiently chiral systems, and nonlinear chirality in the X-ray regime can be enabled by XFELs. With seeded XFELs, high repetition rate, few-femtosecond time resolution and advanced detection schemes, studies of ultrafast dynamics in chiral systems with site specificity yield promising potential for new perspectives on exploring chiral dynamics.

\section{Methods}
\label{sec:methods}	

\subsection{Theoretical model}
\label{sec:methodsTHEORY}

The molecular structure, i.e. assignment of states, their potential energy curves, and the respective vertical ionization potentials were modeled with the recently developed toolkit XMOLECULE \cite{Inhester16,Yajiang19,Inhester19}. To identify the signal from various fragmentation channels, we have computed core-ionization potentials for the neutral molecule and a number of plausible atomic and molecular fragments of different charge. The calculations are based on the $\Delta$-SCF method, where the binding energy is obtained from independent self-consistent-field (SCF) calculation for the initial and final electronic states. We have employed the Hartree-Fock or restricted open-shell Hartree-Fock method, respectively. Specific open-shell electronic configurations, as for example in atomic fluorine, have been taken into account via respective  multi-configuration self-consistent-field (MCSCF) calculations. All calculations have been performed with the 6-311G(d,p) basis set \cite{Krishnan80}. Auger decay rates and core-hole lifetimes have been calculated using the method described in earlier work \cite{Inhester16}. The presently estimated binding energies of the F 1s electrons of different species are in a reasonable agreement with the present observation, providing reliable estimates for the photoelectron energies in Fig.~\ref{fig:Specs}. \\

Calculations of the photoelectron angular distributions were carried out by the single center (SC) method and code \cite{SC1,SC2}, as described in detail in our previous work on trifluoromethyloxirane \cite{Ilchen17}. In particular, calculations were performed in the frozen-core Hartree-Fock approximation at different internuclear geometries between the F$^+$ and  C$_3$H$_3$F$_2$O$^+$ ionic fragments. For numerical convenience, the molecular center was chosen on the carbon atom of the CF$_2$-group of the mother-fragment, to enable an accurate SC-representation of the 1s orbitals of the remaining fluorine atoms. In the calculations, we ignored the 1s core electrons of a dissociating fluorine ion and reduced its nuclear charges accordingly by 2. This allowed us to use shorter SC expansions. In particular, using $\ell_c,\vert m_c \vert < 80$ was sufficient to accurately describe the included 2s and 2p orbitals of a dissociating fluorine ion at the largest considered separation of $\Delta R=14$~a.u. For the photoelectron in the continuum, we used shorter SC expansions with $\ell_\varepsilon,\vert m_\varepsilon \vert < 30$. The working equations required to compute the dichroic parameter are reported in Refs.~\cite{Ilchen17,Knie14}.

\subsection{LCLS machine operation}
\label{sec:methodsLCLS}	

The LCLS provided ultraintense, circularly polarized femtosecond pulses via a periodic, magnetic dipole structure, the so-called Delta undulator, at 120 Hz repetition rate \cite{Lutman16,Hartmann16}. In general, the radiation from an FEL is generated via relativistically moving electrons that are periodically displaced by magnetic chicane arrays, so-called undulators. In contrast to undulators used in storage-ring synchrotron radiation sources, segmented FEL undulators are longer and provide sufficient interaction of the electrons with the emitted light in order to form regions of minimal and maximal localization probability within the electron bunch (coherent microbunching). The light emitted from these fs-substructures is exponentially amplified in subsequent undulator segments.\\

For the present experiment, the LCLS was set up to generate two-color X-ray pulses, one linearly polarized and one circularly polarized. Each pulse was produced in the undulator line by a different electron bunch slice by using the fresh-slice technique \cite{Lutman16b}. This technique is based on impressing a temporal-transverse correlation to the electron bunch upstream of the undulator line and controlling the electron bunch trajectory in two consecutive undulator sections. The linearly polarized X-ray pulse was produced in the first 8 undulator segments from the bunch tail, while the circularly polarized pulse was produced in the variable-polarization Delta undulator from the bunch head, microbunched in the preceding last 8 regular undulator segments and the beam diverting technique \cite{Lutman16, MacArthur18}.\\

In the second section, the first 7 undulator segments were used to microbunch the electrons while the Delta undulator was set to produce circularly polarized X-rays from the microbunched beam. The section used to microbunch the electron beam was pointed downwards, so that the unwanted linearly polarized light generated by the first 7 undulator segments was pointed off the experiment and collimated by a photon collimator. The electron bunch orbit corrector located just upstream of the Delta undulator integrated in a quadrupole magnet was used to kick the bunch upwards \cite{MacArthur18}, to direct the circularly polarized beam to the same location of the first X-ray pulse.\\

Downstream of the first undulator section, a magnetic chicane was used to delay the electrons, thus granting control on the delay between the X-ray photon pulses. When the chicane was turned off, the circularly polarized pulse had a time advance of about 10 to 20 fs on the linearly polarized one, which would correspond to a negative time-delay. By activating the chicane, the circularly polarized pulse could be delayed by up to 1 ps with a resolution of single femtoseconds. The photon energy of each pulse was controlled independently by setting the magnetic strength, \textit{K}, of the undulator segments in the first and the second section at two different values. Switching between right- and left-handed circular polarization is routinely possible within a few minutes by setting the magnetic-pole phase shift of the Delta undulator segment accordingly. The anticipated degree of polarization for these pulses is $>95\%$.

\begin{figure}
\includegraphics[width=0.5\columnwidth]{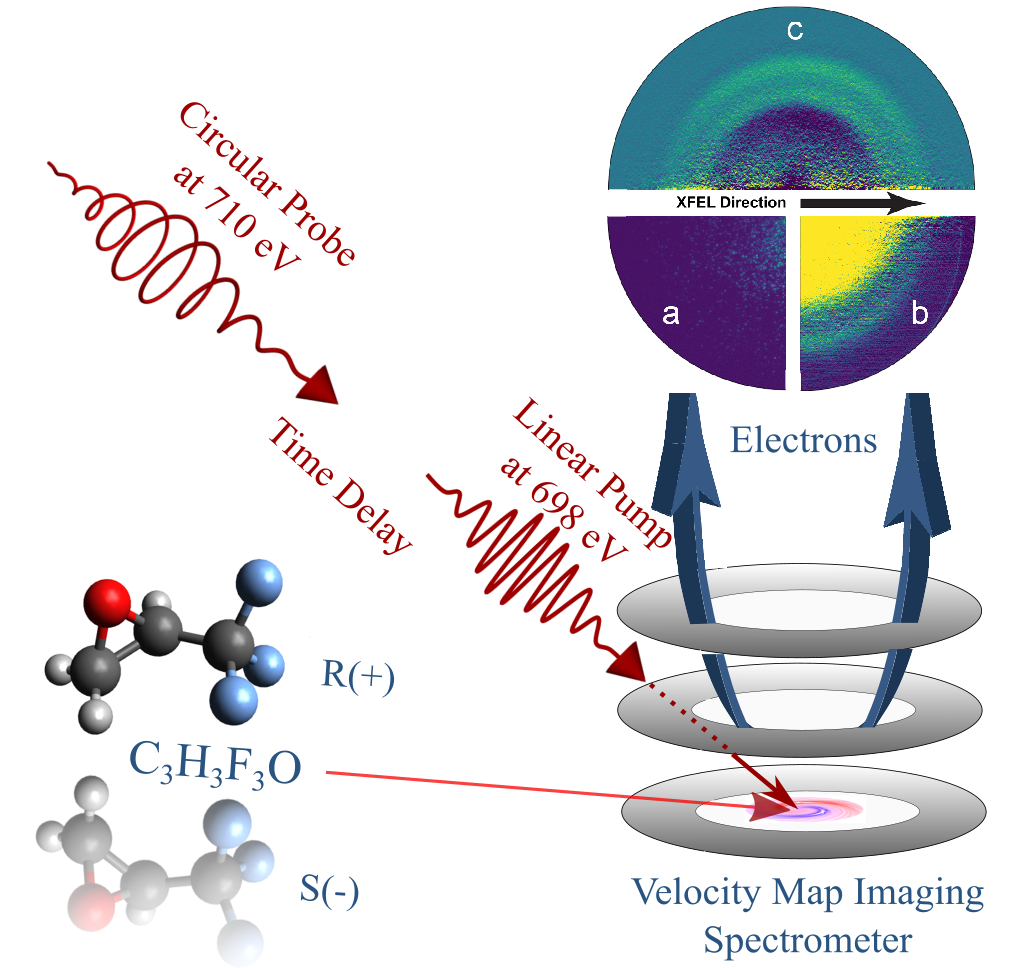}
\caption{Scheme of the experiment. A linearly polarized X-ray pulse triggers the photolysis of C$_3$H$_3$F$_3$O, while a circularly polarized X-ray pulse probes the PECD of 1s electrons emitted from the remaining fluorine-site in the C$_3$H$_3$F$_2$O$^+$ fragment. A VMI spectrometer measures the electrons from the effusively injected molecules with angular resolution. The depicted electron VMI images show one exemplary raw single shot in quadrant 3 (a), an inverted averaged raw image containing information from 150.000 shots in quadrant 2 (b) which is dominated by one-photon processes and basically just reflects the ionization of the neutral molecule by the pump and the probe pulse, and in the upper half (c) the nonlinear signal from the probe pulse, see also Fig. \ref{fig:Specs} for a corresponding angle integrated view. The signal from the singly charged C$_3$H$_3$F$_2$O$^+$ fragment [see contribution c) in Fig.~\ref{fig:Specs}] is in the inner one of the two visible rings, the outer one stems from neutral fluorine atoms [contribution b) in Fig.~\ref{fig:Specs}]. All data corresponds to $\tau$=60 fs.}\label{fig:expt}
\end{figure}

\subsection{Experimental parameters and setup}
\label{sec:methodsSETUP}	

Fig. \ref{fig:expt} illustrates the principal scheme of the experiment. The first-arriving, linearly polarized, pump pulse with h$\nu$=698$\pm$2~eV provided an average pulse energy of $\approx$150 $\mu$J, whereas the second, circularly polarized, probe pulse with h$\nu$=710$\pm$2 eV provided $\approx$50 $\mu$J. The photon energy bandwidth as well as its jitter broaden the linear pump signal to about 5~eV and to about 3.5~eV for the probe pulse. Both pulses had a duration of 10$\pm$3 fs and were focused to 20$\pm$5 $\mu$m FWHM via Kirkpatrick-Baez optics into the LAMP endstation at the AMO beamline of the LCLS \cite{Bozek09, Ferguson15, Osipov18}. With a beamline transmission of 60$\%$ at the chosen photon energies, the irradiation from these pulses in the interaction region with the randomly aligned molecular target was about 2$\times 10^{15}$ W/cm$^{2}$ and 7$\times 10^{14}$ W/cm$^{2}$, respectively, for the pump and probe pulses. This level of irradiation yields a non-negligible level of multiple ionizations within each of these pulses, which complicates the data interpretation and requires a comparison between high and low irradiation levels (see Sec~\ref{sec:methodsDATA}).\\

(R)-(+)+3,3,3-Trifluoro-1,2-methyloxirane and its enantiomeric partner (S)-(--)+3,3,3-Trifluoro-1,2-methyl\-oxirane, each with a purity of 97$\%$ (Sigma Aldrich), were chosen to provide a molecular compound that yields observer sites that are located outside of the stereocenter of the molecule, i.e.\ the fluorine atoms (see molecular representation in Fig.~\ref{fig:expt}). It is experimentally simple to handle in terms of effusive gas-phase delivery due to its vapor pressure of $\approx$530 hPa at room temperature. The three fluorine atoms increase the absolute photoionization cross section at the chosen photon energy to a total of 1.2~Mb which is beneficial for a proof-of-principal experiment, however, it further complicates the differentiation of the expected PECD, since electrons ionized from the individual fluorine atoms have previously been shown to yield very different scattering properties in the molecule \cite{Ilchen17}. This is further substantiated by the theoretical results of Sec.~\ref{sec:ResultsTHEO}. For a detailed interpretation of the data, it furthermore needs to be taken into account that the cross sections of the other elements of the molecule are non-negligible at the chosen photon energies. In particular, the three carbon atoms as well as the single oxygen atom are accounting for $\approx$350 kbarn each, whereas the three hydrogen atoms only contribute $\approx$0.1 kbarn in total. All of these constituents may also undergo multiple photon absorptions at high intensities.\\

Once ejected from the chiral target, the photoelectrons are projected onto a position-sensitive micro-channel-plate (MCP) detector via electrostatic electrodes, i.e.\ a VMI-spectrometer \cite{Eppink97} (see a schematic representation in Fig.~\ref{fig:expt}). The current from the MCPs illuminates a fast phosphorescent screen, i.e.\ type P47, that is read out by a 1 Megapixel low-noise camera. The kinetic energy window defined by the electrostatic field was set to 0--20 eV in order to provide maximum resolution for low kinetic electron energies. The resolution is estimated to be in the order of E/$\Delta$E$>100$. Due to these settings, no Auger electrons were recorded since their kinetic energies lie mainly in between 570 and 650~eV, as calculated by the XMOLECULE toolkit \cite{Inhester16}. Exemplary VMI images for a single shot, an average image as well as the derived nonlinear spectral contributions are depicted in Fig. \ref{fig:expt}.

\subsection{Data interpretation and analysis}
\label{sec:methodsDATA}	

\begin{figure}
\includegraphics[width=0.5\columnwidth]{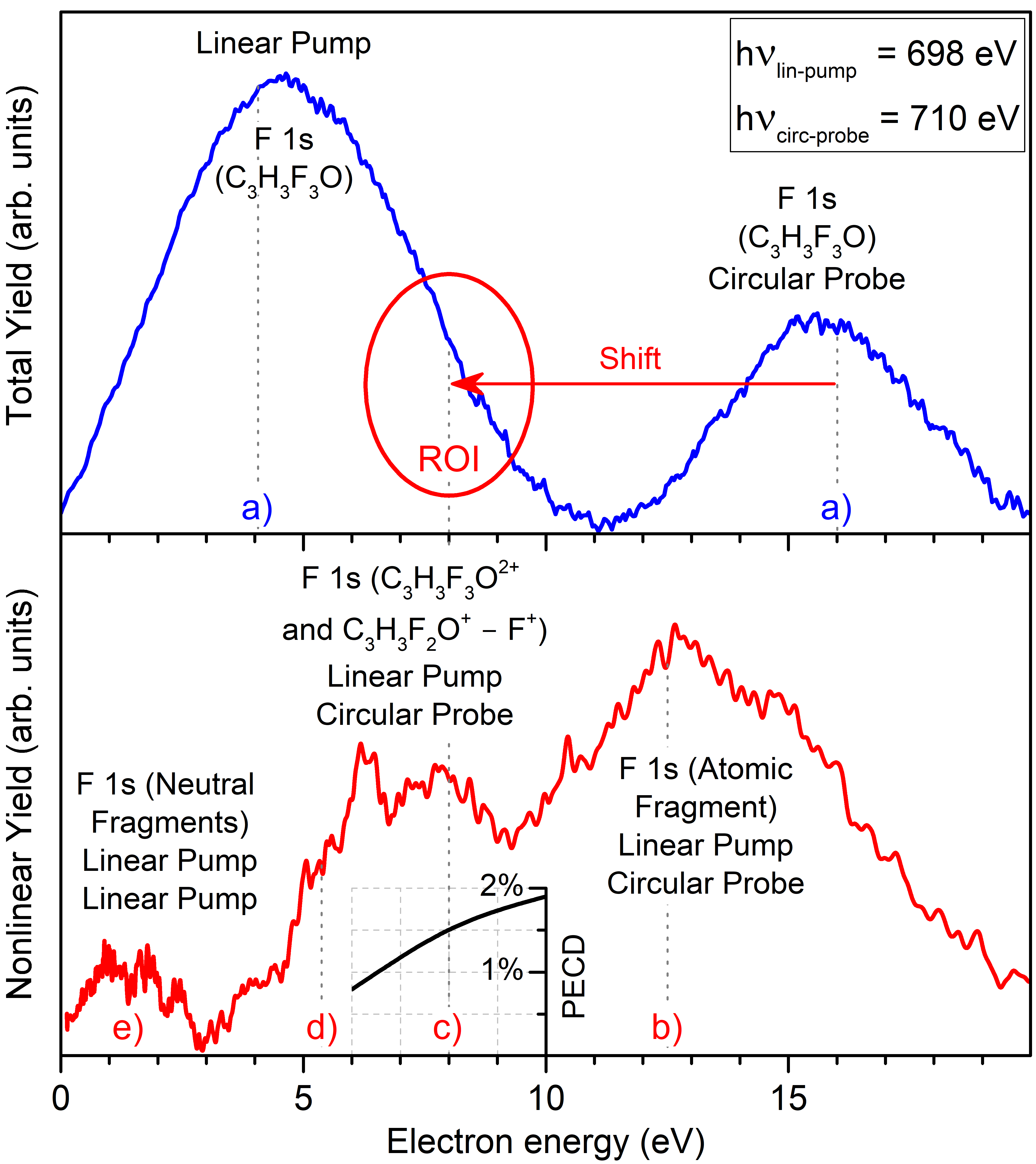}
\caption{Illustration of energy distributions of the total (upper panel) and nonlinear (lower panel) electron emission yields for the exemplary case of $\tau$=60 fs. The former is dominated by the one-photon absorption signals, while the latter by the two-photon absorption signals (as indicated in each panel). The calculated contributions of core electrons of fluorine under different conditions are indicated by letters a)--e) and, for the subject of this study c), an arrow further illustrates the shift. Those linear and nonlinear contributions correspond to: a) one photon ionization of the neutral molecule with the pump or probe pulse, and b)--e) ionization of F 1s in different fragments, as indicated in the panel (see also text for details). An inset to the lower panel illustrates the theoretically predicted kinetic-energy-dependence of the total PECD, computed for the respective internuclear separation of $\Delta R=4$~a.u.}\label{fig:Specs}
\end{figure}

For each combination of polarization, target, and delay, we acquired in average 150,000 images. Resulting from the above described experimental conditions and present theoretical estimates for the F 1s binding energies of the neutral molecule, electrons from one-photon absorption arrived with kinetic energies of about 4~eV, if ionized by the pump pulse, and about 16~eV, if ionized by the probe pulse (see upper panel of Fig.~\ref{fig:Specs}). Those correspond to an F 1s binding energy in the C$_3$H$_3$F$_3$O molecule of $E_{\mathrm{bind}}=694$~eV and are indicated in the upper panel as signals a). As discussed above and in agreement with the experimental data, neither of them are expected to yield a significant PECD. Since only a small fraction of the target ionized by the probe pulse corresponds to the `pumped' target, the one-photon contribution a) was subtracted by filtering the nonlinear pulse-energy progression via sorting for high contributions in the region of interest (more details below). The quadratic progression for the signal over the FEL pulse energy was cross checked, and clearly supports the nonlinear character of the spectrum  highlighted in red in the lower panel of Fig.~\ref{fig:Specs}. \\

As indicated in the upper panel of Fig.~\ref{fig:Specs} with a red arrow, absorption of the first photon followed by the Auger decay and Coulomb explosion of the doubly-ionized molecule results in a shift of the F~1s binding energy for the C$_3$H$_3$F$_2$O$^+$ mother-fragment by about 8~eV, i.e.\ it has a binding energy of $E_{\mathrm{bind}}=702$~eV. The partner-fragment of this dissociation channel, i.e.\ the F$^+$ ion, has a calculated 1s binding energy of 726.6~eV and can therefore not be ionized by any of the pulses.  The probe-pulse will not only release the desired F 1s photoelectrons from the singly-charged mother-fragment C$_3$H$_3$F$_2$O$^+$ [contribution c) in the lower panel of Fig.~\ref{fig:Specs}].  The lower panel of Fig.~\ref{fig:Specs} indicates the presence of further prominent fragments in the `pumped' target, namely: b) the ionization of neutral F atoms by the probe pulse ($E_{\mathrm{bind}}=697.6$~eV), as well as residuals from non-optimal background subtraction of a); d) ionization of the doubly-charged (intact) molecules C$_3$H$_3$F$_3$O$^{2+}$  by the probe pulse ($E_{\mathrm{bind}}=704.7$~eV); and e) signal stemming from sequential ionization by the pump pulse of all neutral fluorine-containing fragments with binding energies in between the neutral molecule and neutral atomic fluorine. Further (unmarked) nonlinear contributions from the probe-pulse could be originated by CF$_3^+$ ($E_{\mathrm{bind}}=704.5$~eV),  CF$_2^+$ ($E_{\mathrm{bind}}=704.9$~eV), CF$^+$ ($E_{\mathrm{bind}}=706.4$~eV) and C$_3$H$_3$F$_2$O$^{2+}$ ($E_{\mathrm{bind}}=707.8$~eV). As mentioned in the previous section, about one third of all fragmenting channels can be originated by the photon-absorption of other atoms in the molecule, which yields very different dynamics and dissociation products that cannot be accounted for in the current experiment.\\

The F 1s photoelectrons of dissociated neutral fluorine atoms account for a significant nonlinear pump-probe signal at a kinetic energy of 12.4~eV. It is interesting to note that despite the large photon energy deposited in the molecule, a significant amount of neutral fluorine atoms is apparently created. This channel is in competition with c) since the underlying processes exclude each other. This signal was analyzed for PECD but, as anticipated due to the large kinetic energy, no sizable PECD was found. 
Contribution d) represents the ionization of the doubly-charged intact molecule that can stem mainly from sequential ionization via the probe-pulse, or at time-overlap, it could occur as combination of photons from the pump and probe pulse.
A small contribution from sequential two-photon absorption from the linearly polarized pump pulse of neutral fragments could explain the signal shifted towards lower kinetic energies $<3$ eV. In the case of a fast dissociation, neutral fluorine atoms can also be ionized by the pump pulse (calculated kinetic energy is about 0.4 eV). Given its bandwidth of about 5 eV, this could in principle contribute to the kinetic energy region below 3~eV. \\

The photon energy jitter can be corrected in the analysis, however, the broad bandwidth inevitably leads to an averaging over the highly energy-dependent PECD. The spatial and timing jitter are negligible due to the chosen operation mode as described in Sec.~\ref{sec:methodsLCLS}. A relative pixel-to-energy calibration of the VMI images was performed by shifting the photon energy of the pump pulse between 690~eV and 715~eV. The absolute photon energy is derived from the calculated binding energy of the neutral molecule. Due to the complex machine setup and a resulting lack of time, a common transmission-normalization to the racemic target was only done for 60~fs delay. However, due to the subtraction of normalized images from different targets and different polarizations under otherwise similar conditions, the PECD determination for the other time delays was still possible.\\

After normalization and calibration of the VMI-images, they are sorted into different intensity regimes of the incoming light as well as signal strength in the ROI (see Fig.~\ref{fig:Specs}). This is feasible due to the stochastic nature of the photon generation, which causes a large variation of the pulse energy from shot to shot. It can be regarded as intrinsic intensity scan. The images were subsequently Abel-inverted \cite{Vrakking01} in order to recover the original three-dimensional distribution of the electron emission. Due to a small damaged area on the detector in one quadrant, only data from one half of the detector was taken into account. Subtracting the average of the lower half of the intensity contributions in the ROI from the average of the highest ten percent, as identified to be the best trade-off between strong pump-probe signal and low contributions from sequential ionization, allowed for self-referencing of the individual spectra. This was cross checked for consistency with a dedicated background subtraction of individual runs containing no circular color. In Fig. \ref{fig:expt}, the pump-probe data is furthermore illustrated as VMI image (upper half). As reference, also averaged data from the highest 10\% of the recorded pulse energies (lower right) as well as an exemplary single shot are depicted.

\newpage

For deriving the PECD, the such processed normalized images acquired for left- and right-handed circular polarizations for the same enantiomer are subtracted from each other. The resulting asymmetric signal in the full three-dimensional angular distribution for the observed F 1s region of C$_3$H$_3$F$_2$O$^+$ was then quantified for its forward--backward asymmetry percentage. Contributions b), c), and d) were fit with Gaussian curves in order to extract the individual contribution of c). Due to this angle and energy integration of signal c) for individual quadrants, it is not feasible to derive an energy-resolved PECD. For 60 fs, we have additionally confirmed the anticipated sign-change of the forward--backward asymmetry with swapping from R(+) to S(-) enantiomers. For 250~fs delay, this method of enantiomer-change was the only available data set, which is part of the reason for the largest of the uncertainties. We have partly encountered a slight energy and bandwidth shift after swapping helicities of the photons, which also results in a larger uncertainty due to slightly different kinetic energy contributions to the PECD. Combining these imperfections with the estimated additional uncertainties of the manifold processes after inner-shell ionization by the pump pulse, the complex machine conditions, the partly damaged detector and therefore reduced statistics, the broad bandwidth of the photons and their jitter as well as the challenging derivation of the nonlinear signal and its angular distribution, causes the overall experimental uncertainty to be large (see Fig.~\ref{fig:results}). Therefore, only the 60 fs delay reveals an unambiguous PECD. It should, however, be stressed that the robust probe signal of the circularly polarized pulse, as shown in the lower part of Fig.~\ref{fig:Specs}, is a significant outcome of the experiment in itself. The present experimental procedure does not provide access to the energy-resolved PECD of signal c). An inset to the lower panel of Fig.~\ref{fig:Specs} illustrates that the averaged total PECD, computed for the internuclear separation of $\Delta R=4$~a.u. corresponding to the time delay of 60~fs, increases smoothly from about 0.8\% to 1.9\% in the kinetic energy range of 6--10~eV, yielding the energy-averaged value of about 1.4\%.\\

\begin{acknowledgements}
We gratefully acknowledge the fruitful discussions with M. Larsson (University of Stockholm, Sweden), V. Zhaunerchyk (University of Gothenburg, Sweden), B. Erk (DESY, Germany), L. Nahon (SOLEIL, France), and R. Boll (European XFEL, Germany). Use of the Linac Coherent Light Source (LCLS), SLAC National Accelerator Laboratory, is supported by the U.S. Department of Energy, Office of Science, Office of Basic Energy Sciences under Contract No. DE-AC02-76SF00515. The authors acknowledge the invaluable support of the technical and scientific staff of the LCLS at SLAC National Accelerator Laboratory. This work was funded in part by the Deutsche Forschungsgemeinschaft (DFG) -- Projektnummer 328961117 -- SFB 1319 ELCH (Extreme light for sensing and driving molecular chirality). M.I., P.S., V.M., and Z.L. acknowledge funding from the Volkswagen Foundation within a Peter Paul Ewald-Fellowship. G.H. and A.K. acknowledge funding from the BMBF (05K16RKA) and A.O.L. from the Knut and Alice Wallenberg Foundation. P.R. acknowledges funding from the Studienstiftung des deutschen Volkes and from the German Academic Exchange Service DAAD. S.B. was supported by the Helmholtz Initiative and Networking Fund through the Young Investigators Group Program and acknowledges support from the Deutsche Forschungsgemeinschaft, project B03/SFB755. T.J.A.W. is supported by the U.S. Department of Energy, Office of Science, Basic Energy Sciences, Chemical Sciences, Geosciences, and Biosciences Division.
\end{acknowledgements}

\vspace{0.5cm}
\textbf{Author contributions}\\
M.I., A.K., and S.M. have conceived the experiment.
The experiment was performed by M.I., Ph.S., G.H., P.R., R.N.C., A.G., N.H., W.H., T.O., D.R., P.W., T.W., S.B., A.K., and S.M.
Machine operation and diagnostics were conducted by A.A.L., J.P.M., T.M., H.-D.N., and Z.H.
P.S., G.H., and M.I. have analyzed the data.
N.M.N., L.I., Z.L., and Ph.V.D. performed the theoretical calculations.
M.I. and Ph.V.D. have written the paper with dedicated contributions by Z.L., L.I., A.A.L., and further contributions from all co-authors.


\begin{thebibliography}{99}

\bibitem{Ritchie76}
B. Ritchie, \textit{Theory of the Angular Distribution of Photoelectrons Ejected from Optically Active Molecules and Molecular Negative Ions}, Phys. Rev. A \textbf{13}, 1411 (1976).

\bibitem{Nahon15}
L. Nahon, G. A. Garcia, and I. Powis, \textit{Valence Shell One-Photon Photoelectron Circular Dichroism in Chiral Systems}, J. Electron Spectrosc. Relat. Phenom.  \textbf{204}, 322 (2015).

\bibitem{Hadidi18} R. Hadidi, D. K. Bozanic, G. A. Garcia, and L. Nahon, \textit{Electron Asymmetries in the Photoionization of Chiral Molecules: Possible Astrophysical Implications}, Adv. Phys. X \textbf{3}, 1477530 (2018).

\bibitem{Hergenhahn04}
U. Hergenhahn, E. E. Rennie, O. Kugeler, S. Marburger, T. Lischke, I. Powis, and G. Garcia, \textit{Photoelectron Circular Dichroism in Core Level Ionization of Randomly Oriented Pure Enantiomers of the Chiral Molecule Camphor}, J. Chem. Phys. \textbf{120}, 4553 (2004).

\bibitem{Ilchen17}
M. Ilchen, G. Hartmann, P. Rupprecht, A. N. Artemyev, R. N. Coffee, Z. Li, H. Ohldag, H. Ogasawara, T. Osipov, D. Ray, \textit{et al.}, \textit{Emitter-Site-Selective Photoelectron Circular Dichroism of Trifluoromethyloxirane},  Phys. Rev. A \textbf{95}, 053423 (2017).

\bibitem{Lux12}
C. Lux, M. Wollenhaupt, T. Bolze, Q. Liang, J. Kohler, C. Sarpe, and T. Baumert, \textit{Circular Dichroism in the Photoelectron Angular Distributions of Camphor and Fenchone from Multiphoton Ionization with Femtosecond Laser Pulses}, Angew. Chem. Int. Ed. \textbf{51}, 5001 (2012).

\bibitem{Lehmann13}
C. S. Lehmann, N. B. Ram, I. Powis, and M. H. M. Janssen, \textit{Imaging Photoelectron Circular Dichroism of Chiral Molecules by Femtosecond Multiphoton Coincidence Detection}, J. Chem. Phys. \textbf{139}, 234307 (2013).

\bibitem{Beaulieu16}
S. Beaulieu, S. Comby, B. Fabre, D. Descamps, A. Ferr\'{e}, G. Garcia, R. G\'{e}neaux, F. L\'{e}gar\'{e}, L. Nahon, S. Petit, \textit{et al.}, \textit{Probing Ultrafast Dynamics of Chiral Molecules Using Time-Resolved Photoelectron Circular Dichroism}, Faraday Discuss. \textbf{194}, 325 (2016).

\bibitem{Comby16}	
A. Comby, S. Beaulieu, M. Boggio-Pasqua, D. Descamps, F. L\'{e}gare\'{e}, L. Nahon, S. Petit, B. Pons, B. Fabre, Y. Mairesse, \textit{et al.}, \textit{Relaxation Dynamics in Photoexcited Chiral Molecules Studied by Time-Resolved Photoelectron Circular Dichroism: Toward Chiral Femtochemistry}, J. Phys. Chem. Lett. \textbf{7}, 4514 (2016).

\bibitem{Wollenhauptreview}
M. Wollenhaupt, \textit{Universality of Photoelectron Circular Dichroism in the Photoionization of Chiral Molecules}, New J. Phys. \textbf{18}, 102002 (2016).

\bibitem{optXray}
B. Erk, R. Boll, S. Trippel, D. Anielski, L. Foucar, B. Rudek, S. W. Epp, R. Coffee, S. Carron, S. Schorb, \textit{et al.}, \textit{Imaging Charge Transfer in Iodomethane Upon X-Ray Photoabsorption}, Science \textbf{345}, 288 (2014).

\bibitem{XrayXray1}
K. Schnorr, A. Senftleben, M. Kurka, A. Rudenko, G. Schmid, T. Pfeifer, K. Meyer, M. K\"{u}bel, M. F. Kling, Y. H. Jiang, \textit{et al.}, \textit{Electron Rearrangement Dynamics in Dissociating I$_2^{n+}$ Molecules Accessed by Extreme Ultraviolet Pump-Probe Experiments}, Phys. Rev. Lett. \textbf{113}, 073001 (2014).

\bibitem{XrayXray2}
C. E. Liekhus-Schmaltz, I. Tenney, T, Osipov, A. Sanchez-Gonzalez, N. Berrah, R. Boll, C. Bomme, C. Bostedt, J. D. Bozek, S. Carron, \textit{et al.}, \textit{Ultrafast Isomerization Initiated by X-Ray Core Ionization}, Nat. Commun. \textbf{6}, 8199 (2015).

\bibitem{XrayXray3}
K. R. Ferguson, M. Bucher, T. Gorkhover, S. Boutet, H. Fukuzawa, J. E. Koglin, Y. Kumagai, A. A. Lutman, A. Marinelli, M. Messerschmidt, \textit{et al.}, \textit{Transient Lattice Contraction in the Solid-to-Plasma Transition}, Sci. Adv. \textbf{2}, e1500837 (2016).

\bibitem{XrayXray4}
C. S. Lehmann, A. Pic\'{o}n, C. Bostedt, A. Rudenko, A. Marinelli, D. Moonshiram, T. Osipov, D. Rolles, N. Berrah, C. Bomme, \textit{et al.}, \textit{Ultrafast X-Ray-Induced Nuclear Dynamics in Diatomic Molecules Using Femtosecond X-Ray-Pump-X-Ray-Probe Spectroscopy}, Phys. Rev. A \textbf{94}, 013426 (2016).

\bibitem{XrayXray5}
A. Pic\'{o}n, C. S. Lehmann, C. Bostedt, A. Rudenko, A. Marinelli, T. Osipov, D. Rolles, N. Berrah, C. Bomme, M. Bucher, \textit{et al.}, \textit{Hetero-Site-Specific X-Ray Pump-Probe Spectroscopy for Femtosecond Intramolecular Dynamics}, Nat. Commun. \textbf{7}, 11652 (2016).

\bibitem{XrayXray6}
N. L. Opara, I. Mohacsi, M. Makita, D. Castano-Diez, A. Diaz, P. Jurani\'{c}, M. Marsh, A. Meents, C. J. Milne, A. Mozzanica, \textit{et al.}, \textit{Demonstration of Femtosecond X-Ray Pump X-Ray Probe Diffraction on Protein Crystals}, Struct. Dynam. \textbf{5}, 054303 (2018).

\bibitem{Allaria14}
E. Allaria, B. Diviacco, C. Callegari, P. Finetti, B. Mahieu, J. Viefhaus, M. Zangrando, G. De Ninno, G. Lambert, E. Ferrari, \textit{et al.}, \textit{Control of the Polarization of a Vacuum-Ultraviolet, High-Gain, Free-Electron Laser}, Phys. Rev. X \textbf{4}, 041040 (2014).

\bibitem{Mazza14}
T. Mazza, M. Ilchen, A. J. Rafipoor, C. Callegari, P. Finetti, O. Plekan, K. C. Prince, R. Richter, M. B. Danailov, A. Demidovich, \textit{et al.}, \textit{Determining the Polarization State of an Extreme Ultraviolet Free-Electron Laser Beam Using Atomic Circular Dichroism}, Nat. Commun. \textbf{5}, 4648 (2014).

\bibitem{Lutman16}
A. A. Lutman, J. P. MacArthur, M. Ilchen, A. O. Lindahl, J. Buck, R. N. Coffee, G. L. Dakovski, L. Dammann, Y. Ding, H. A. D\"{u}rr, \textit{et al.}, \textit{Polarization Control in an X-Ray Free-Electron Laser}, Nat. Photon. \textbf{10}, 468 (2016).

\bibitem{Hartmann16}
G. Hartmann, A. O. Lindahl, A. Knie, N. Hartmann, A. A. Lutman, J. P. MacArthur, I. Shevchuk, J. Buck, A. Galler, J. M. Glownia, \textit{et al.}, \textit{Circular Dichroism Measurements at an X-Ray Free-Electron Laser With Polarization Control}, Rev. Sci. Instrum. \textbf{87}, 083113 (2016).

\bibitem{Iwayama13}
H. Iwayama, N. Sisourat, P. Lablanquie, F. Penent, J. Palaudoux, L. Andric, J. H. D. Eland, K. Bu\v{c}ar, M. \v{Z}itnik, Y. Velkov, \textit{et al.}, \textit{A Local Chemical Environment Effect in Site-Specific Auger Spectra of Ethyl Trifluoroacetate}, J. Chem. Phys. \textbf{138}, 024306 (2013).

\bibitem{SC1}
Ph. V. Demekhin, A. Ehresmann, V. L. Sukhorukov, \textit{Single Center Method: A Computational Tool for Ionization and Electronic Excitation Studies of Molecules}, J. Chem. Phys. \textbf{134}, 024113 (2011).

\bibitem{SC2}
S. A. Galitskiy, A. N. Artemyev, K. J\"{a}nk\"{a}l\"{a}, B. M. Lagutin, and Ph. V. Demekhin, \textit{Hartree-Fock Calculation of the Differential Photoionization Cross Sections of Small Li Clusters}, J. Chem. Phys. \textbf{142}, 034306 (2015).

\bibitem{LCLS_AMO}
Webpage of the LCLS Beamline AMO:\\ https://lcls.slac.stanford.edu/instruments/amo

\bibitem{Woerner19}
D. Baykusheva, D. Zindel, V. Svoboda, E. Bommeli, M. Ochsner, A. Tehlar, H. J. W\"{o}rner, \textit{Real-Time Probing of Chirality During a Chemical Reaction}, Proc. Natl. Acad. Sci. USA,  \textbf{116}, 23923 (2019).

\bibitem{Inhester16}
L. Inhester, K. Hanasaki, Y. Hao, S.-K. Son, and R. Santra, \textit{X-Ray Multiphoton Ionization Dynamics of a Water Molecule Irradiated by an x-Ray Free-Electron Laser Pulse}, Phys. Rev. A \textbf{94}, 023422 (2016).

\bibitem{Yajiang19}
Y. Hao, L. Inhester, S.-K. Son, and R. Santra, \textit{Theoretical Evidence for the Sensitivity of Charge-Rearrangement-Enhanced x-Ray Ionization to Molecular Size}, Phys. Rev. A \textbf{100}, 013402 (2019).

\bibitem{Inhester19}
L. Inhester, Z. Li,  X. Zhu, N. Medvedev, and T. J. A. Wolf, \textit{Spectroscopic Signature of Chemical Bond Dissociation Revealed by Calculated Core-Electron Spectra}, J. Phys. Chem. Lett. \textbf{10} 6536 (2019).

\bibitem{Krishnan80} 
R. Krishnan, J.S. Binkley, R. Seeger, and J. A. Pople, \textit{Self‐consistent Molecular Orbital Methods. XX. A Basis Set for Correlated Wave Functions}, J. Chem. Phys.  \textbf{72}, 650 (1980).

\bibitem{Knie14}
A. Knie, M. Ilchen, P. Schmidt, P. Rei{\ss}, C. Ozga, B. Kambs, A. Hans, N. M\"{u}glich, S. A. Galitskiy, L. Glaser, \textit{et al.}, \textit{Angle-Resolved Study of Resonant Auger Decay and Fluorescence Emission Processes After Core Excitations of the Terminal and Central Nitrogen Atoms in $N_2O$}, Phys. Rev. A \textbf{90}, 013416 (2014).

\bibitem{Lutman16b}
A. A. Lutman, T. J. Maxwell, J. P. MacArthur, M. W. Guetg, N. Berrah, R. N. Coffee, Y. Ding, Z. Huang, A. Marinelli, S. Moeller, \textit{et al.}, \textit{Fresh-Slice Multicolour X-Ray Free-Electron Lasers}, Nat. Photon. \textbf{10}, 745 (2016).

\bibitem{MacArthur18}
J. P. MacArthur, A. A. Lutman, J. Krzywinski, and Z. Huang, \textit{Microbunch Rotation and Coherent Undulator Radiation from a Kicked Electron Beam}, Phys. Rev. X \textbf{8}, 041036 (2018).

\bibitem{Bozek09}
J. D. Bozek, \textit{AMO Instrumentation for the LCLS X-Ray FEL}, Eur. Phys. J. - Special Topics \textbf{169}, 129 (2009).

\bibitem{Ferguson15}
K. R. Ferguson, M. Bucher, J. D. Bozek, S. Carron, J.-C. Castagna, R. Coffee,  G. I. Curiel, M. Holmes, J. Krzywinski, M. Messerschmidt, \textit{et al.}, \textit{The Atomic, Molecular and Optical Science Instrument at the Linac Coherent Light Source}, J. Synchr. Rad. \textbf{22}, 492 (2015).

\bibitem{Osipov18}
T. Osipov, C. Bostedt, J.-C. Castagna, K. R. Ferguson, M. Bucher, S. C. Montero, M. L. Swiggers, R. Obaid, D. Rolles, A. Rudenko, \textit{et al.}, \textit{The LAMP Instrument at the Linac Coherent Light Source Free-Electron Laser}, Rev. Sci. Instrum. \textbf{89}, 035112 (2018).

\bibitem{Eppink97}
A. T. J. B. Eppink and D. H. Parker, \textit{Velocity Map Imaging of Ions and Electrons Using Electrostatic Lenses: Application in Photoelectron and Photofragment Ion Imaging of Molecular Oxygen}, Rev. Sci. Instrum. \textbf{68}, 347 (1997).

\bibitem{Vrakking01}
M. J. J. Vrakking, \textit{An Iterative Procedure for the Inversion of Two-Dimensional Ion/Photoelectron Imaging Experiments}, Rev. Sci. Instrum.  \textbf{72}, 4084 (2001).


\end{thebibliography}
\end{document}